# Marine Integrated Energy Microgrids

Chendan Li
*Department of Marine Technology*
Norwegian University of Science and Technology
Trondheim, Norway
chendan.li@ntnu.no

Muzaidi Othman
*Faculty of Electrical Engineering Technology,*
University Malaysia Perlis (UniMAP)
Perlis, Malaysia
muzaidi@unimap.edu.my

Nor Baizura Ahamad
*Faculty of Electrical Engineering Technology,*
University Malaysia Perlis (UniMAP)
Perlis, Malaysia
baizura@unimap.edu.my

Marta Molinas
*Department of Engineering Cybernetics*
Norwegian University of Science and Technology
Trondheim, Norway
marta.molinas@ntnu.no

*Abstract*—Marine sector decarbonization is another important battlefield for meeting the goal of climate action and ensuring the fulfillment of ambitions for a zero-emission society. Driven immediately by the policy incentives such as Energy Efficiency Design Index (EEDI) from International Maritime Organization(IMO), carbon taxation and labeling, a series of innovations centered around marine transportation are emerging from both industry and academia. As an efficient energy system form, the microgrid is playing an increasingly important role as the system constitution form for various marine energy systems. With specific concern about multi-energy integrations, the conventional definition of the microgrid needs also to be extended for an integrated energy system. In this paper, we will first introduce the extended concept of the microgrid as an integrated energy system and its applications in the marine sector, and then present the state of the art for the control, operation, and system integration for it and its clusters. The challenges and opportunities for marine integrated energy microgrids will also be discussed to shed light on future research.

*Keywords—Maritime, microgrid, integrated energy systems*

## I. Introduction

The decarbonated future is electric. Electrification as an important means for efficiency improvement can significantly reduce the greenhouse gas(GHG) emissions in the marine sector, especially for $CO_2$, where global shipping accounts for around 3% of global energy-related $CO_2$. Electrification brings more precise control and lower losses from energy conversion and transmission and in total a higher round-trip efficiency to many traditional marine energy systems, such as the traditional fossil fuel based mechanical propulsion system, diesel engine based drilling platform, etc. Another drive for electrification is the increasing electricity demand from the ocean activities coming from more advanced electrical facilities, machines, and devices. Technology maturity, regulatory and environmental appeals are adding more momentum to this positive transition driving the entire marine sector towards sustainability. One concrete example is that IMO proposes an EEDI index for the design stage of the ship, in order to achieve a 50% GHG reduction by 2050 from international shipping [1].

As an efficient energy system form, the autonomous electrical power network, microgrid, can be found in many marine applications. There are several previous works discussing marine microgrids on shipboard [2], on the seaport[3], on remote islands, etc. As the fuel economy of shipping is shifting from the current domination of mineral oils, to cleaner fuels such as biofuels, liquid natural gas (LNG), ethane, methanol, and totally zero emission solutions such as hydrogen, fully electric, and nuclear, with its viability depending on different ship types and business evolved, and some largely driven by the regulation, multiple energy sources find their niche in the various marine hybrid energy systems. In this sense, the integrated energy system with the electrical network as its key part can be taken as an integrated energy microgrid (IEM).

In this paper, the concept of IEM is borrowed to cope with multiple energy carriers for marine applications extending the conventional definition of the microgrid. The paper is organized as follows. After introducing the concept of the IEM, in Section II, the marine applications that fall into the category of IEM will be listed. In Section III, the state of the art of control, operation, and system integration will be introduced for these IEMs with the focus on onboard IEM and its extension to the entire marine integrated energy system. The challenges of developing these IEMs and opportunities in terms of the emerging research topics will be reviewed in Section IV. Finally, Section V gives the conclusion.

## II. The Concept of Integrated Energy Microgrid(IEM) and Its Marine Applications

According to the new IEEE Std 2030.7-2017, for a system to be considered as a microgrid, it must have: a) clearly defined electrical boundaries. b) a control system to manage and dispatch resources as a single controllable entity. c) installed generation capacity that exceeds the critical load; this allows the microgrid to be disconnected from the main grid, i.e., operate as an entity in islanded mode, and supply the local loads[4]. Following this norm, we can make a similar definition for the IEM to extend the concept of microgrids to integrated energy systems. An IEM can thus be defined as a multi-energy network with a clearly defined boundary that can supply its critical energy loads with the control of its own dedicated energy management controller[5].

The marine energy system is a dynamic field that includes many of these IEMs. Fig. 1 shows the examples of these marine IEMs ranked by the power ratings. Following is the introduction of each of these typical IEMs.

*1) Shipboard IEM*
Maritime logistics have undertaken more than 80% of the global trading demand since the 1970s, and electrical energy is becoming the dominating energy form to power the propulsion, communication, navigation, and other subsystems on a vessel through the power network despite ship types. Yet, two factors make the shipboard as an IEM necessary. One factor is the path of electrification necessitates the diverse energy storage/primary energy source to provide the electricity, which varies from carbon neutral fuels to various energy storages. Another factor is that the load on the ship, especially for passenger ships, includes forms that are more than just electricity, such as heating and cooling. Typical shipboard IEMs include ships with hybrid propulsion systems (usually deepsea ships), fully electric ships with alternative fuel as primary energy other than electric batteries (which use the fuel cell to convert the primary energy source to electricity), and ships with none-electricity loads.

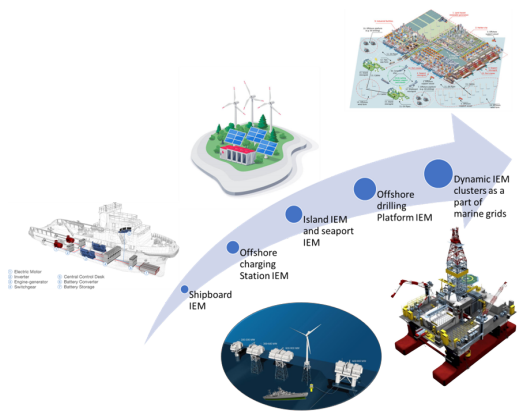

Fig. 1. Examples of marine integrated energy microgrids

2)   Offshore charging station as IEM

As the energy capacity and specific energy of most of the fossil fuel free energy sources, such as battery and ammonia, are still not comparable with the traditional fuel, charging in the middle of the voyage becomes a necessity for the more electrified and low carbon shipping. Unlike the ship powered by fossil fuel which is refilled by bunker barges coming to them, they need a charging station to provide the electricity. This need spawns the application of offshore charging stations. For example, the project Stillstrom established by Maersk Supply and Orsted will launch a first-to-market product in offshore charging, allowing idle vessels to use renewable electricity by connecting to the charging buoys [6]. Besides directly recharging with electricity, it can be estimated that other energy networks such as hydrogen might also be included for these offshore charging station applications in the short future. With different energy carriers, load and generation, future offshore charging stations are promising IEMs.

3)   Remote island as IEM

Microgrid is among the first choices for the utility to provide electricity to rural places and small islands where it is not economically feasible to be connected with the main utility grid. With the need to lower the emission, renewables are increasingly adopted other than diesel generators in these microgrids. The energy storage becomes decisive for a valid business case, where the medium to long term storage capability necessitates the inclusion of other energy carriers such as hydrogen and its necessary conversion systems. One of the examples is the EU project REMOTE with several pilots demonstrating power-to-gas and gas-to-power systems [7].

4)   Seaport IEM

Environmental issues such as $CO_2$ emissions, air pollution, noise, vessel congestion, and waste (sewage and solid) are the main problems that plague the traditional port. Improving energy efficiency by integrating more renewables and deepening electrification to form a seaport microgrid become the solution to promote automation as well as alleviate these pollution problems. With multiple alternative fuels for the ships and variety of the energy storages to meet the new need of the green transition, these seaport microgrids evolve into integrated energy systems, such as Groningen seaports of Amsterdam with the abundant wind power to form the hydrogen ecosystems.

5)   Offshore drilling platform as IEM

Offshore drilling utilizes petroleum reserves beneath the Earth's oceans instead of those on land. To respond to the pledge of decarbonization, powering the rigs with renewable energy is gaining more attention. There is a new hybrid energy system for offshore oil and gas installations that presents itself in [8]. Offshore wind power, on-site gas turbines, and a fuel cell and electrolyze stack energy storage device are all part of the design. With this mix of energy sources, different kinds of drilling platforms including drilling ships become IEMs that interact with offshore power plants and the seaport IEM.

6)   An integrated marine energy system with IEM clusters

IEMs located in the same or closed sea areas are not isolated systems. Upgraded and emerging drilling platforms need alternative energy sources other than fossil fuel to operate and offshore service vessels to conduct routine maintenance activities; more electrified ships need to charge either while berthing at the seaport or at the offshore charging stations; modern seaport and low carbon islands will involve more electrified loads and other energy sources to supply facilities with more human activities. All these IEMs form a more complex system of systems featured as the IEM clusters. The coordination of all the energy carriers and network optimization requires more sophisticated engineering such as multi-disciplinary design optimization, multi-carrier flow analysis, interdependency analysis, communication network design, information system interoperation, etc, and is worth further research

III.   CONTROL, OPERATION AND SYSTEM INTEGRATION OF MARINE IEMS

A.   *Shipboard IEM–State of the Art of Its Electrical Part*

A ship is a complex engineering system. Its design covers the naval architecture design, steady stability and hydrodynamic design, seakeeping and maneuvering design for the hull, as well as the onboard power system design and other activities such as project management to fulfill various functions. These interactive, iterative and/or integrated design processes are aided with computer aided design/manufacture/engineering softwares to synthesize these multidisciplinary domains [9]. Electrical part of the whole system, which is often taken as a microgrid, is the key part to make the many subsystems such as HVAC, dynamic positioning system, powertrain, communication system function properly, and which might be a part of the onboard IEM. It paves the foundation to drive the modern ship marching toward a more connected, more autonomous, and more electrified system.

The development of the shipboard microgrid is in line with the evolution of the onboard propulsion system. It evolves from introducing the genset to the purely machinery system by combining the prime mover such as engines and the alternator to serve the motor for propelling, to the fully electric propulsion system in which more advanced power electronics drive interfaces and motors supplied by the power directly from various electricity energy storage and generation system, such as batteries, fuel cell, flywheels, supercapacitor to small nuclear reactors. The architecture of this onboard microgrid evolves from a purely AC network, to hybrid AC/DC networks, to currently the widely recognized solutions, i.e., purely DC network which suits the onboard system electrification with small volume, high efficiency, and lower power losses. The ring topology is increasingly adopted instead of the radial for reliability purposes, and to adapt to the system protection, the topology design will divide the network into zones.

Hierarchical control and operations is a proven system design methodology for not only power systems but also other systems, such as TCP/IP for the internet and computer communication networks. This is also the control and operational method for the electrical part of the shipboard microgrid. In general, the lower layer control is mostly for the internal and faster control tasks, the upper layers deal more with the global optimization which does not only care about control within the system but also how it interacts with the external systems, with a relatively slower control loop. For example, the primary control could deal with the parallel operation of homogeneous generators and energy storages, as well as the coordination between homogenous energy storages, such as that between batteries and supercapacitors. The secondary control of the electrical shipboard microgrid often deals with the mitigation of the power quality issues, such as voltage/frequency deviations and harmonics, as well as the operation mode alteration[10]. The upper layer control provides the optimization for the power dispatch to achieve multiple objectives, such as emission reductions, cost reductions, and lifetime extensions. etc[11]. It is also the interface between the shipboard microgrid and the system outside the vessel. For example, the emerging autonomous ships will be required to connect to a remote operation center, the upper layer function will be moved to this remote operation center, and even the engine control room function can be conducted remotely in this remote operation center.

For the control approach, the literature often sees three different types, which are centralized, decentralized, and distributed control. Centralized control has only one controller, i.e., centralized controller, while the other two have more than one separated controller, achieving the global objective either through communication (for distributed control ) or without communication (decentralized control). With the development of ICT technology, increased connectivity is witnessed for the control and operations of the shipboard which is characterized by the feature of moving part of their control functionality to the cloud, to conduct more sophisticated computations.

The protection of the shipboard microgrid is a crucial part of system safety, especially true for purely DC networks and the fully electric system which contains several new energy sources such as batteries or hydrogen storage which have higher safety requirements. For DC networks, particular standards are established for not only recommending voltage levels but also the recommended system architecture and protection method, such as [12], [13]. Besides, standards dedicated to shipboard power systems will also shed light on the protection, which include IEC 63108, IEC 61660, IEC 60092-507, IEC 61892-1, and IEEE 1709.

The main difference between the onboard microgrid and that on land is that the system weight and space are often important design parameters for the first, despite these two having different mission profiles and ambient working conditions. For example, the shipboard microgrid might have a higher Peak to Average Power Ratio (PAPR), more frequent and higher magnitude load change, and even regenerative load. Therefore, it incurs several special concerns for the design of the electrical system. The integrated design which will connect hull design, layout design, and powertrain design will be a promising direction. Hybrid energy storage that balances the energy density and power density is also a necessary solution being underinvested now.

The increasing application of the power electronic system and advanced electric machines brings higher power density and smaller volume for a more optimized electric/electrical (E/E) architecture, and more stringent requirements on the reliability of the system. On one hand, high power density motors, such as permanent magnetic motors, semi-superconducting, and superconducting motors, are reducing the powertrain volume with high requirements for cooling systems. On the other hand, new Wide Band Gap (WBG) technologies such as Gallium Nitride (GaN) and Silicon Carbide (SiC) devices require more research on their reliability, especially regarding their production process and other factors that influence the system performances such as EMC and thermal management.

*B. Energy Integration for Shipboard IEMs*

Improved system reliability, efficiency, and reduced emissions are all possible with the help of multiple energy sources on the supply side, thanks to the IEM concept applied to the ships. The supply side of energy can be categorized as the prime mover, the energy storage system (ESS), and the renewable energy source. Prime movers include sources like diesel engines, gas turbines, and fuel cells. Prime mover in an Integrated Power System (IPS) for instance, is dominated by diesel generators (DG) which produce emissions such as $CO_2$, $NO_x$, $SO_x$, and particulate matter. Fuel cell and energy storage systems (ESS), such as batteries, are the cleaner options for the supply side. ESS in ships has a function to compensate for the power fluctuations caused by shiploads and intermittencies of renewable energy. For more effective power fluctuation compensation, different ESS technologies can be combined to achieve better energy and power density characteristics.

Current research demonstrates the promising combination of Li-ion batteries and supercapacitors.

The evolving alternative fuel pathway necessitates various fuel networks to be combined with the onboard power system, especially for ships with limited upgrading. To reduce GHG emissions, eco-friendly fuels have been used as alternative fuels in diesel generators. For example, using Liquefied Petroleum Gas (LPG), methanol, ethanol, ammonia, and hydrogen have been considered [15]. Hydrogen is one of the clean energy carriers that can be relied on in the long run; however, due to its flammability and the cryogenic conditions when stored as liquid hydrogen, ensuring its safety presents a significant challenge [16]. To comply with the Emission Control Areas (ECAs), a dual fuel diesel generator system seems to be suitable as it can be switched between different fuels subject to availability and cost. It combines diesel fuel and natural gas such as LNG. Ammonia has good chemical and thermodynamic properties as an alternative fuel, but its ignition requirements such as low flame speed, high minimum ignition energy, high auto-ignition temperature, and narrow flammability limits make it difficult to use as single fuel without mixing with other fuels or decomposition [16]. Other methods to reduce GHG are In-Cylinder Combustion and Combustion Post-Treatment as mentioned in [17].

Integration with the carbon network with the shipboard IEM is also possible. A carbon Capture System (CCS) can be considered for the ship since it also contributes to the reduction of GHG which is mainly used in land-based generators. CCS captures and stores $CO_2$ in a gas tank. However, the system integration will become complicated, with high investment costs, and require larger space onboard. In addition, this will increase the source's power consumption. Thus, to prevent power shortages, it could be properly sized while taking into account the power consumption from ESS [18].

Integration also comes from the load side. The heat from waste heat recovery can be used to provide valuable energies either in the form of power or thermal and increase overall efficiency. Large cruise ships, for example, have a huge amount of usage of heat and cooling energy demand for the hotel facilities. The heat energy demand can be utilized from the boiler which converts from the heat energy waste rather than operating the oil-fired boiler [19]. For instance, energy conversion from chemical energy, namely fuel-cell (e.g.: SOFC and PEMFC) to electrical, which is considered suitable for marine application and produces quite a low emission. Thus, heat from the high-temperature exhaust from SOFC and PEMFC which are around 800°C and 70°C respectively, and Micro Gas Turbines (mGTs) around 280°C can be used for thermal energy recovery systems [20]. In [21], FC's waste heat recovery applies a Recuperative Organic Rankine Cycle (RORC) system to produce power and other uses for heat recovery for refrigeration cycle condensers. In addition to reusing heat for thermal recovery, GHG emissions such as ($CO_2$, $SO_x$, $NO_x$) can be selectively captured and stored in a storage tank. This is accomplished via Carbon Capture System (CCS) and could aid in reducing GHG emissions up to 70%.

Fig. 2 show the possible topology of a IEM for cruise ship with integration of several energy flows and loads such as DG, Fuel Cell, Combined Cooling/Heating power (CCHP) generator, ESS(e.g.: Battery, Flywheel, Supercapacitor), Renewable Energy Source (e.g.: PV, Wind Turbine), electrical load side (e.g.: Propulsion load, Ship Service Load, Power to Cooling/Heating), thermal load side, thermal storage, and CCS.

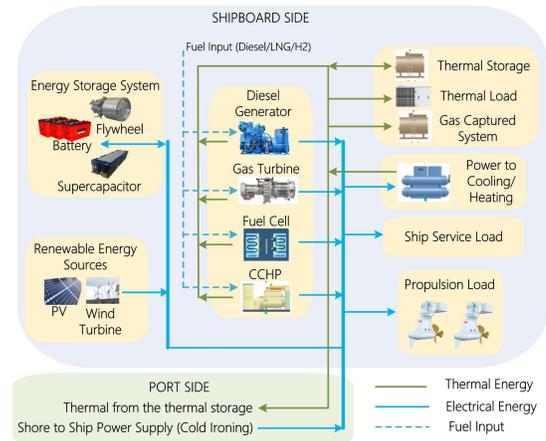

Fig. 2. Possible Topology of an IEM for a cruise ship with different energy storage systems, loads and Gas/Carbon Captured System.

### C. System Coordination of Multiple IEMs as IEM Clusters

The shipboard IEM is a special case of IEMs which shares many similarities with other IEMs. The key is to efficiently integrate multiple energy systems to achieve overall optimality. For instance, when a ship berths at port, it connects to the seaport IEM. With increasing electricity charging demand for electrified vessels, besides the grid capacity enhancement method such as installing batteries on the port, coordinating charging of the electric vessels will also provide the large power capacity needed. As renewables will be the energy resources to support the port coming either from the land or offshore, it requires the charging to coordinate with these intermittent generations as well as the local load from the land where the port is located, with direct incentives from the electricity market. The shipping routing planning and scheduling, together with the operation of the oil platform can also be integrated to achieve higher efficiency. Similar cases would apply to the planning and operation of other energy networks, such as hydrogen and LNG which provides the primary energy to the mobile IEMs — ships, across the sea. Each IEM will behave as an energy hub [22], coordination and optimization will make this cluster utilize the multiple energy flows more optimally while meeting the demand.

## IV. CHALLENGES AND OPPORTUNITIES

Marine integrated energy microgrid as a problem solver merges multiple energy networks and requires multidisciplinary knowledge. Several hurdles are still there to be overcome and require continuous research and development effort. Essentially, the technical challenges come from the increased system complexity and multidisciplinary nature of the system. To design a more efficient, more optimized, and demand-Oriented IEM, an integrated design methodology and tools are needed. For the shipboard IEM, inspired by control co-design for ship design, the IEM system design should also not only combine the multiple energy flow analysis but also connect with the engineering of the whole system, such as the fluid dynamic of the vessel, structure, etc, to achieve a more optimized system holistically. Extending the system integration possibilities become increasingly important to boost the connectivity of multiple IEMs, which can witness more system integration applications in parallel with IEMs, such as the shipping schedule and routing planning being connected with the IEM operations. For achieving this, the digital application to merge multiple information systems and data analysis will be the key enabler.

Increased system complexity also brings more high-level challenges such as safety, security, and reliability. Cryogenic fuels such as LNG, and liquid hydrogen bring more challenges to safety, which requires new technology not only for their usage but also for the testing and auditing of their safety permanence. Moreover, increased adoption of ICT and system integration increases the interdependency of subsystems and enlarges the attack surface in terms of cybersecurity, especially for the emerging autonomous shipping systems. To quantify and to mitigate the risk for a more integrated and connected system becomes an urgent topic. The challenges and opportunities listed above among many other difficulties need additional efforts to fill the knowledge gaps where system engineering awareness and competence are of paramount importance.

## V. Conclusion

In this paper, we have introduced the emerging paradigm of marine integrated energy microgrids. We first introduced the example of the marine IEMs, and later introduced the key technological state of the art for these systems, where projecting that marine energy systems will form a cluster of IEMs for the marine energy system. Challenges and opportunities to utilize these systems better are also given which highlight the research directions urgently needed. In future work, we will improve our review on the existing example and technologies by elaborating the state of the art, and investigating novel methods for the planning, operation, and control of marine IEMs.